# Carrier separation in type II quantum dots inserted in (Zn,Mg)Te/ZnSe nanowire

P Baranowski[1], M. Szymura[1], M. Wójcik[1], R. Georgiev[2], S. Chusnutdinow[1], G. Karczewski[1], T. Wojtowicz[3], L.T. Baczewski[1] and P. Wojnar[1]

[1] *Institute of Physics, Polish Academy of Sciences, 02-668 Warsaw, Poland*
[2] *Institute of Optical Materials and Technologies "Acad. J. Malinowski", Bulgarian Academy of Sciences, Akad. G. Bonchev str., bl. 109, 1113 Sofia, Bulgaria.*
[3] *International Research Centre MagTop, Institute of Physics, Polish Academy of Sciences, 02-668 Warsaw, Poland*

**Abstract**

Quantum dots consisting of an axial $Zn_{0.97}Mg_{0.03}Te$ insertion inside a large bandgap $Zn_{0.9}Mg_{0.1}Te$ nanowire cores are fabricated in a molecular beam epitaxy system by employing the vapor-liquid-solid growth mechanism. Additionally, this structure is coated with a thin ZnSe radial shell which forms type II interface with the dot semiconductor. The resulting radial electron-hole separation is evidenced by several distinct effects which occur in the presence of ZnSe shell, including: the optical emission redshift of about 250 meV, a significant decrease of the emission intensity, the increase of the excitonic lifetime by one order of magnitude and the increase of the biexciton binding energy. The type II nanowire quantum dots where electrons and holes are radially separated constitute a promising platform for potential applications in the field of quantum information technology.

**Introduction**

Semiconductor nanowires belong to the most intensively studied nanostructures in the last decade. They have drawn an interest as the promising platform for nanoelectronics[1,2], light sources[3,4] and light detectors[5,6]. Characteristic feature of these nanostructures is the greater flexibility to combine semiconductors with different lattice constants as compared to their planar counterparts. Moreover, different semiconductors can be merged either in the radial or axial direction depending on the growth direction. Radial heterostructures have been used as building blocks of solar cells[7], nano-lasers[8] or light emitting photodiodes[9], whereas axial heterostructures are investigated mainly in view of their possible applications for single photon emission[10,11]. The latter effects are observed in nanowire quantum dots (NWQD) i.e., small optically active axial insertions of a low band gap semiconductor inside a nanowire composed of a large bandgap semiconductor forming zero-dimensional traps for electrons and holes[12–14].

The heterostructures can be classified either as type-I or type-II, or type III depending on the relative bandgap offsets. In particular, type-II heterostructures are characterized by a staggered bandgap alignment, which means that the energies of both the conduction and valence band edges of one semiconductor are shifted towards the higher energy with respect of the corresponding band edges of the other semiconductor. As an effect, electrons and holes tend to transfer through the interface in the opposite directions, resulting in the spatial separation of charge carriers and subsequent built-in electric field. An efficient control of the carriers separation and transportation determines the performance of the optoelectronic devices, such as solar cells[15–17] or photodetectors[18–20]. Moreover, the interfacial transition in type II heterostructures enables the spectral range extension to significantly longer wavelength defined by the bandgaps of semiconductor components and relative band offsets[21–23].

In this study, we present the optical properties of molecular beam epitaxy grown type II NWQD in which the electrons and holes are separated in the radial direction. In the axial direction the nanowire core is composed of $Zn_{0.9}Mg_{0.1}Te$ and contains a quantum dot built of $Zn_{0.97}Mg_{0.03}Te$, i.e., a semiconductor with a smaller band gap. In the absence of radial ZnSe shell, both electrons and holes localize within this part of the nanowire. However, the presence of the ZnSe shell which forms a type II interface with $Zn_{0.97}Mg_{0.03}Te$ dots leads the electrons to be pushed outside of the dot forming a ring shaped wavefunction around the dot, provided that no additional electron localization occurs due to, e.g., shell thickness inhomogeneity. The delocalization of the electron wavefunction within the entire ZnSe shell along the NW axis is prevented by the attractive electron-hole Coulomb interaction. A scheme of the investigated structure is presented in Figure 1a.

The intentional design of the band structure in the heterostructures presented in this work enables the effective control over the charge carriers wavefunctions on the nanometer scale which may open a path towards novel applications. In particular, type II NWQD with radial electron-hole separation could be particularly well suited for the observation of the excitonic Aharonov-Bohm effect. This quantum interference effect rises an interest due to its potential applications in the field of quantum information storage[24–26]. There is only a limited number of publications reporting the optical Aharonov-Bohm effect in various structures including self assembled quantum rings[27,28], quantum tubes[29], type-II QDs columns[30]. Quite recently, this effect has been also observed in a nanowire heterostructure where the carrier separation takes place at type II wurtzite/zinc blende crystal phase interfaces[31]. Here we report a new concept of the intentional type II NWQD growth in which the electrons-holes are separated in the radial direction. The large difference between the electron and hole radiuses should lead to the decrease of the magnetic field required for the observation of Aharonov-Bohm oscillations which may be a great advantage of the NW geometry described in this paper.

**Growth**

Nanowire heterostructures investigated in this work are grown in the molecular beam epitaxy (MBE) setup consisting of two growth chambers (EPI 620 system and Prevac) coupled by an ultra-high vacuum connection. In the first chamber Te, Zn and Mg atomic fluxes are set to be characterized by the beam equivalent pressure (BEP) equal to $4.5 \times 10^{-7}$, $2 \times 10^{-7}$ and $1.5 \times 10^{-8}$ Torr, respectively. In the second chamber Zn flux is set to $4 \times 10^{-7}$ and Se flux to $2 \times 10^{-7}$ Torr. $Zn_{0.97}Mg_{0.03}Te$ axial quantum dots inside the $Zn_{0.9}Mg_{0.1}Te$ NWs are grown in the first MBE chamber following a procedure described in details previously[32]. In short, the gold-silicon catalyst droplets with the diameters ranging from 20 nm to 50 nm are formed by thermal dewetting of a 1 nm-thick gold layer deposited on (111)-oriented silicon substrate. Then, the vapor-liquid-solid growth of $Zn_{0.9}Mg_{0.1}Te$ NW-cores takes place at 410°C for 23 min. It is followed by the low bandgap segment composed nominally of $Zn_{0.97}Mg_{0.03}Te$ which is grown for 15 s by alternate closing of Mg-flux for 1 s and opening it for 0.5 s. A typical length of the low bandgap segment is estimated to be 15 nm. The NW top segment composed of $Zn_{0.9}Mg_{0.1}Te$ is grown for 3 min and corresponds to the average length of about 130 nm. In the next step, the sample is transferred to the second MBE chamber for ZnSe shell deposition. After stabilizing the temperature at 300°C in the Zn-flux, the sample is exposed to Se-flux for a period of time that is varied in the range from 60 s to 300 s. Finally, nanowires are transferred back to the first MBE chamber for the deposition of $Zn_{0.8}Mg_{0.2}Te$ outer shell which serves as the surface passivation layer, whereas Mg concentration in the shell is estimated based on previously studied reference samples[33]. After stabilizing the substrate temperature at 350°C in the Te flux, the growth of the outer shell takes 10 minutes and leads to an average $Zn_{0.8}Mg_{0.2}Te$ shell thickness of about 15 nm. The scheme of the nanowire heterostructure grown by the above described procedure is shown in Figure 1a.

There are several parameters of the investigated nanowire heterostructure which were subject to an optimization procedure including: the length of the low bandgap $Zn_{0.97}Mg_{0.03}Te$ insertion, Mg concentration within the NW core and within the NWQD, the thickness of ZnSe shell. Our goal was to obtain the most intensive optical emission from $Zn_{0.97}Mg_{0.03}Te$ NWQDs. In particular, it is found that further increase of ZnSe shell thickness results in a disappearance of any optical emission from NWQDs. This effect is caused most likely by an efficient electron-hole separation process taking place at the dot/ZnSe-shell interface.

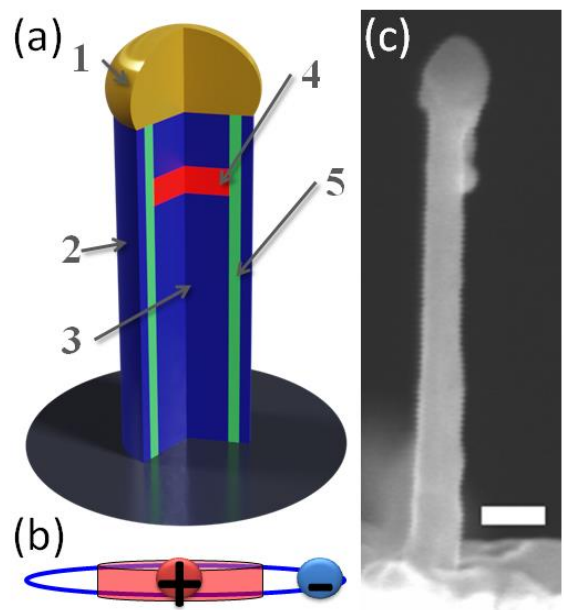

**Figure 1** a) Scheme of investigated nanowire heterostructure 1- Au/Si eutectic droplet, 2- $Zn_{0.8}Mg_{0.2}Te$ outermost shell, 3-$Zn_{0.9}Mg_{0.1}Te$ core, 4- $Zn_{0.97}Mg_{0.03}Te$ low bandgap insertion, 5-ZnSe internal shell. b) Scheme representing spatial separation of electrons and holes occurring at NWQD/ZnSe shell interface. c) Scanning electron microscopy of a typical nanowire heterostructure investigated in this work. Scale bar corresponds to 100 nm.

**Results and discussion**

The NWs morphology is studied by means of scanning electron microscopy (SEM). For these measurements the electron beam acceleration voltage is set to 5 kV and the probe current to 40 pA. In Figure 1c, a side-view of a typical NW is presented. After a study of several NWs, we find that their length varies from 0.8 to 2 μm and their diameter from 70 to 90 nm depending on the wire. Bare $Zn_{0.9}Mg_{0.1}Te$ NWs without shells are significantly thinner and their diameters are in the range of 40 – 60 nm. By comparing these diameters the average thickness of $Zn_{0.8}Mg_{0.2}Te$ outer shell amounts to ~15 nm what is consistent with the values obtained from the calibrated beam fluxes.

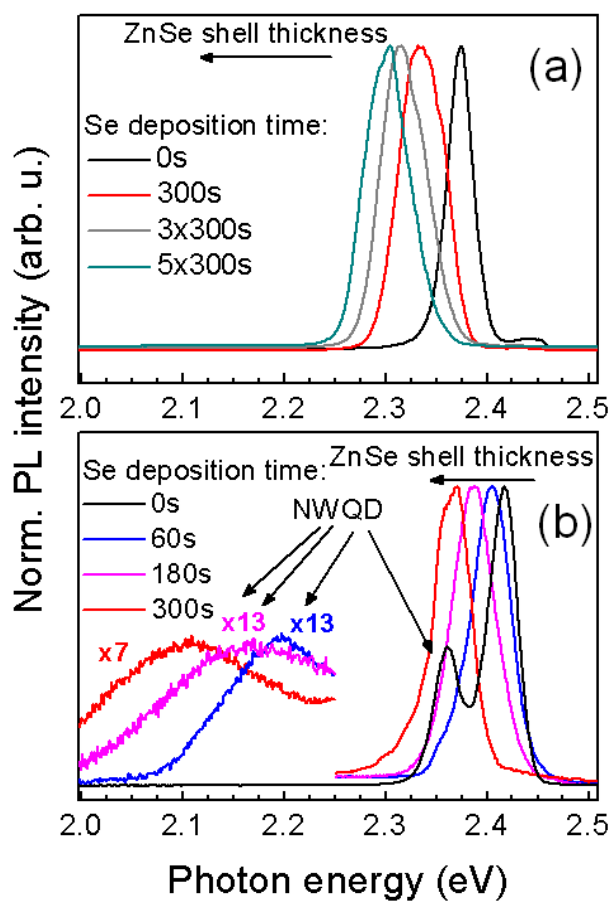

**Figure 2** Normalized photoluminescence (PL) spectra from $Zn_{0.9}Mg_{0.1}Te/ZnSe/Zn_{0.8}Mg_{0.2}Te$ core/shell/shell nanowires with different thickness of ZnSe shell (a) without and (b) with $Zn_{0.97}Mg_{0.03}Te$ axial nanowire quantum dot (NWQD). Black line represents the spectrum from a reference samples without ZnSe shell. A significant redshift of the optical emission from $Zn_{0.9}Mg_{0.1}Te$ core and from NWQDs is observed with increasing ZnSe shell thickness. (a) Se deposition time amounts to 0s, 1x300s, 3x300s, 5x300s for black, red, grey and green lines respectively, and in (b) 0s, 60s, 180s and 300s for black, blue, magenta and red lines, respectively. The temperature of the measurement is 7 K and the excitation wavelength - 405 nm.

The optical characterization of 'as grown' NW heterostructures starts with the low-temperature photoluminescence (PL) measurements. The samples are placed in a closed cycle cryostat at 7 K and are excited with a 405 nm solid state laser. The beam is focused to the spot of 0.1 mm diameter. The signal is detected in a 303 mm monochromator (SR303i by Andor) equipped with a CCD camera. In Figure 2a, the optical emission spectra of $Zn_{0.9}Mg_{0.1}Te/ZnSe/Zn_{0.8}Mg_{0.2}Te$ core/shell/shell NWs without NWQDs are presented as a reference. PL-spectra from four different samples with varying ZnSe shell thickness ranging from zero up to 5 monolayers are presented. The highest emission energy of 2.38 eV is measured for the NWs without ZnSe shell, whereas a pronounced redshift of this energy down to 2.30 eV is observed with an increasing ZnSe shell thickness.

In Figure 2b, the optical emission from $Zn_{0.9}Mg_{0.1}Te/ZnSe/Zn_{0.8}Mg_{0.2}Te$ core/shell/shell NW heterostructures but containing an axial $Zn_{0.97}Mg_{0.03}Te$ NWQD are presented. The presence of this insertion results in the appearance of an additional emission band at energies lower than the emission from NW cores. For instance, the optical emission from the NW heterostructure without ZnSe shell (black spectrum in Figure 2b) consists of two emission lines: the line at 2.40 eV originates from the NW cores and the line at 2.36 eV from NWQDs. With an increasing ZnSe deposition time the NWQD-related emission exhibits also a significant redshift down to 2.11 eV. Simultaneously, the emission intensity drops by about one order of magnitude.

The redshift of the optical emission can be explained by the type II character of the $Zn_{1-x}Mg_xTe/ZnSe$ interface. The low temperature bandgaps of unstrained ZnTe and ZnSe amount to 2.39 eV and 2.82 eV, respectively, whereas the type II interlayer transition takes place at 2.0 eV[34]. Holes are expected to localize within ZnTe and electrons - within ZnSe. When introducing Mg atoms into ZnTe matrix, the interlayer transition is expected to shift of about ~0.02 eV per 1 atomic percent toward higher energies, whereas the type II character of the $Zn_{1-x}Mg_xTe/ZnSe$ interface should be preserved in the entire Mg-concentration range. Increasing the ZnSe shell thickness causes the electron wavefunction to be pushed outside of the NW cores towards ZnSe shell. Then, the probability of confinement of electrons in ZnSe shell increases causing the emission energy decrease toward the values predicted by the $Zn_{1-x}Mg_xTe/ZnSe$ interlayer transition. This interpretation in terms of radial electron-hole separation in the presence of ZnSe shell is consistent with the observation of a significant drop of the optical emission intensity. The latter effect is a direct consequence of the decrease of the electron – hole wavefunction overlap.

The impact of other factors which may influence the emission energy, such as strain and quantum size effect is less important compared to the previously mentioned type II character of the band alignment. The NW-core diameters of 30 – 60 nm are too large for the observation of the quantum size effect. Moreover, ZnSe exerts a compressive strain on ZnTe and $Zn_{0.9}Mg_{0.1}Te$ cores which should result in the strain-induced increase of the emission energy. Instead, a decrease of the emission energy is observed which demonstrates that the strain effect has not a decisive impact on the emission energy in the studied structures.

Based on the results presented in Figure 2, it is not possible to associate definitely the observed PL-bands to the NW-core and NWQD based emission. It is, in principle, possible that any luminescence may originate from other structures grown simultaneously with the proper straight nanowires, such as from two-dimensional residual deposits between NWs or from any crooked nanostructures. In order to identify unambiguously the origin of the optical emission bands cathodoluminescence (CL) measurements are performed. For this purpose Zeiss Ultra55 SEM equipped with a system for CL signal collection and detection (HCLUE by Horiba) and a helium-cooled sample stage (by Kammrath and Weiss) are used. The measurements are performed on individual NWs which are removed from the original substrate and dispersed on a clean Si substrate by using the sonication in an isopropanol bath.

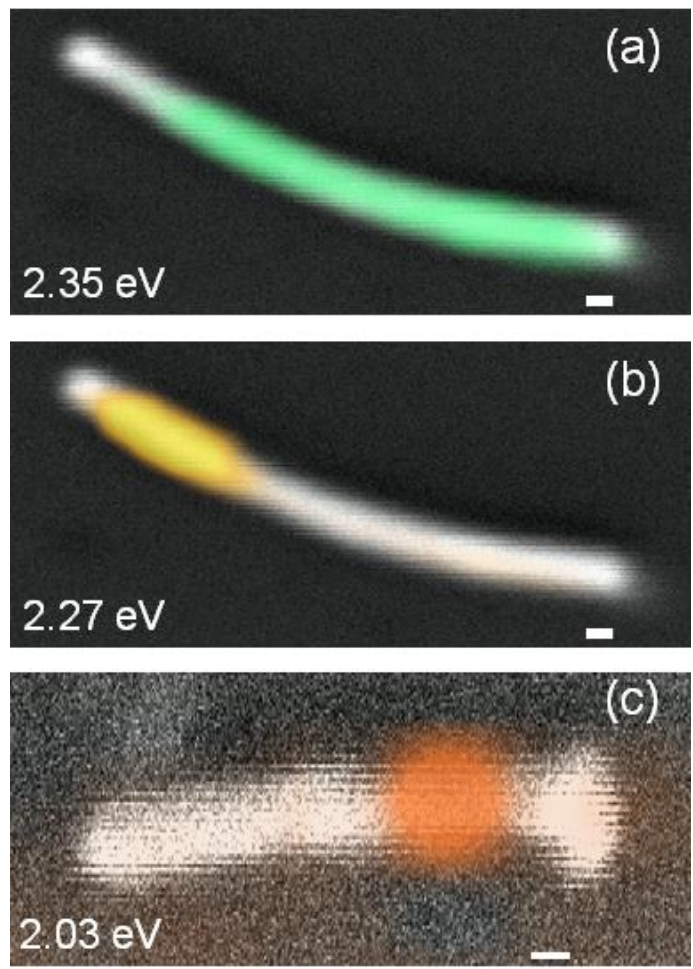

**Figure 3** Cathodoluminescence (CL) maps superimposed with SEM pictures from individual $Zn_{0.9}Mg_{0.1}Te$ nanowires with $Zn_{0.97}Mg_{0.03}Te$ axial insertions and $ZnSe/Zn_{0.8}Mg_{0.2}Te$ radial shells (a) CL map from a structure with a thin ZnSe shell (deposited during 20 s) performed at 2.35 eV demonstrates that this emission originates from the nanowire core and (b) the emission at 2.27 eV comes from the nanowire quantum dot. (c) CL map from a structure with a thick ZnSe shell (deposited during 300 s). The emission at 2.03 eV originated from the type II nanowire quantum dot. All scale bars correspond to 100 nm. Temperature of the measurement is 7 K, electron acceleration voltage – 15 keV and the probe current – 1.2 nA.

Cathodoluminescence from the two different individual $Zn_{0.9}Mg_{0.1}Te$ nanowires with $Zn_{0.97}Mg_{0.03}Te$ axial insertions and $ZnSe/Zn_{0.8}Mg_{0.2}Te$ radial shells are presented in Figure 3. The difference between these two NW-heterostructures is the thickness of ZnSe-shell. Se-deposition time was 20 s and 300 s in the case of the NW presented in Figure 3a-3b and 3c, respectively. CL map performed at 2.35 eV reveals that the emitting area corresponds to the almost entire NW, Figure 3a. This fact leads to the conclusion that it is due to the recombination involving the entire $Zn_{0.9}Mg_{0.1}Te$ core. In contrast, CL-map performed at 2.27 eV, Figure 3b, shows that the emitting area is significantly smaller and placed close to the top of the NW which corresponds to the position of the $Zn_{0.97}Mg_{0.03}Te$ axial insertion (yellow area). In the case of the NW heterostructure with a thicker ZnSe shell presented in Figure 3c, it is demonstrated that the weak optical emission at 2.03 eV comes, indeed, from the type-II NWQD. Similar measurements are performed on various NW heterostructures with different ZnSe shell thicknesses. Most importantly, it is found that all PL-lines shown in Figure 2 originate from the corresponding NW heterostructures. When two emission lines are measured, the high energy line is always due to excitonic recombination within the NW-core and the low energy line comes from the NWQD. The emission energy from NWQDs decreases by a relatively large value of 250 meV with an increasing ZnSe shell thickness.

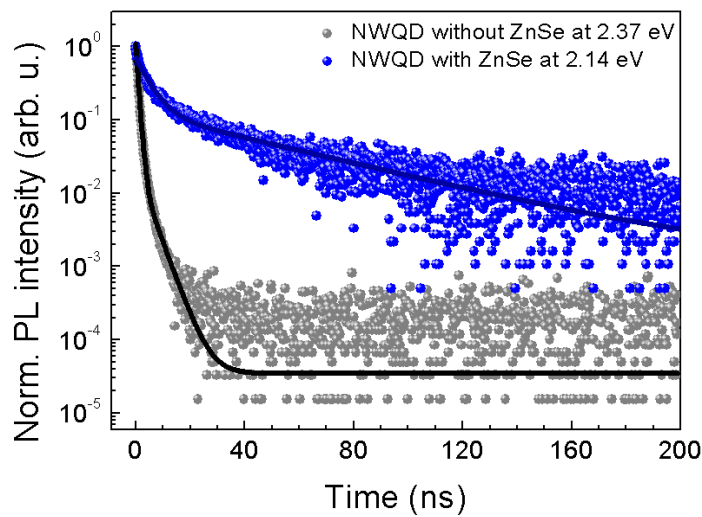

**Figure 4.** Photoluminescence decay from nanowire quantum dots. Gray points come from the reference structure without ZnSe shell in which the dots are characterized by the type I confinement. Blue decay comes from nanowire quantum dots with radial ZnSe shell deposited for 300s. In the latter structure the electron-hole separation is manifested by a significant increase of the decay time. Temperature of the measurements is 7 K and the excitation laser line wavelength - 400 nm.

The electron-hole separation at the $Zn_{0.97}Mg_{0.03}Te/ZnSe$ dot/shell interface is also clearly manifested in time resolved PL (TRPL) measurements. Namely, the reduction of the electron-hole overlap should result in a significant increase of excitonic life-time in type II quantum dots as compared to their type I counterparts. In order to conduct TRPL measurements, samples are excited by picoseconds laser pulses at a wavelength of 405 nm. The temporal pulse width of 128 ps with a repetition rate of 5 MHz is used for TRPL measurements of reference sample without ZnSe internal shell, whereas the same temporal pulse width with a repetition rate of 2 MHz is used for NWQDs with ZnSe internal shell, where significantly longer decays are observed. Single photon counting avalanche photodiode is used for the signal detection.

Normalized TRPL traces from NWQDs with and without ZnSe internal shell are presented in Figure 4. As a reference, the photoluminescence decay is measured from type I NWQDs grown without ZnSe shell which spectrum is presented in Figure 2b as a black line. Only photons originating from the low energy band, i.e., from NWQDs, are considered. It is found that the decay time from this reference sample is close to the temporal resolution of the setup, grey points in Figure 4. The decay time increases significantly for sample with NWs having an additional radial ZnSe shell. In Figure 4, the PL decay from NWQDs with the thickest ZnSe shell is presented (blue points), whereas the corresponding spectrum is shown as a red line in Figure 2b in which the NWQD-related emission band appears at 2.14 eV.

The TRPL-traces from both samples are fitted with a biexponential decay. In type II NWQDs with ZnSe-shells the long decay time is found to be 48 ns whereas the short decay time amounts to 5 ns. In contrast, the characteristic decay time values from type I NWQDs are noticeably shorter and amount to 4 ns and 0.8 ns, respectively. The necessity to use biexponential function for the fits of

PL-decay traces can be explained by the presence of emission lines from single- and multiexcitonic complexes within the ensemble emission from NWQDs. The long decay time corresponds to the single exciton emission whereas the fast component to multiexcitonic complexes, such as biexcitons, in which the relatively large number of decay channels results in an effective reduction of the lifetime.

The most important message from the TRPL study is, however, that the presence of radial ZnSe shell results in a significant increase of the excitonic life time within the NWQDs which is a direct consequence of the spatial electron-hole separation and confirms the presence of type II band alignment in NWQDs with a radial ZnSe shell.

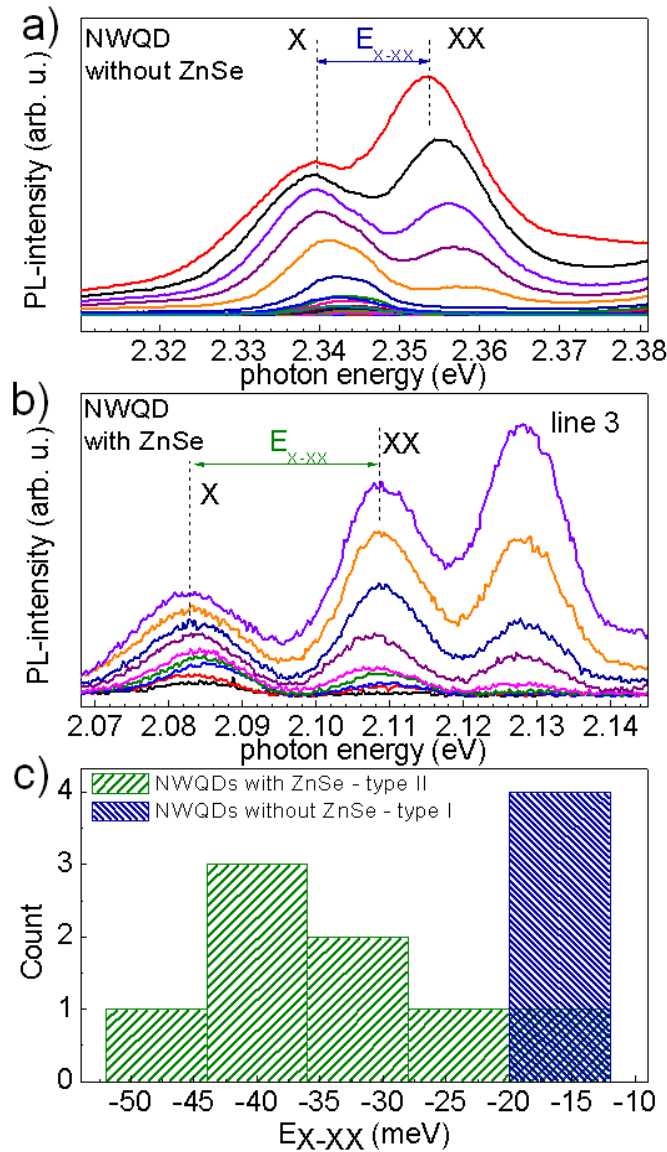

**Figure 5** µ-PL spectra of an individual NWQD with increasing excitation power without (a) and with ZnSe shell (b). Excitonic, X, and biexcitonic, XX, emission can be identified. The temperature of the measurement is 7 K and the excitation wavelength 405 nm. c) Histogram of the biexciton binding energy ($E_X-E_{XX}$) for type II, in green, and type I NWQDs, in blue.

Optical emission from individual NWs has also been studied by means of micro-photoluminescence (µ-PL). The laser beam with the wavelength of 405 nm is focused onto a ~3 µm diameter spot with a microscope objective which allows for addressing the emission from individual NWs. Then, the emitted light is collected by the same objective and detected by a 500 mm-monochromator (SR-

500i by Andor) equipped with a CCD camera. Sample is placed on a cold finger inside a cryostat at 7 K. In order to resolve spatially the emission from individual NWs, the NWs are dispersed on a clean silicon substrate.

First of all, it is found that the broad emission bands shown in Figure 2b which have been previously identified to originate from NWQDs split into several sharp lines with the spectral width of the order of a few meV coming from individual nanostructures. The emission from the particular dot depends on the diameter of the NW, the length of the dot, as well as the thicknesses of both shells which define the strain conditions in the structure. That is why the emission energy from different NWQD may slightly vary depending on the particular dot.

In Figure 5a and 5b the emission spectrum dependence on the excitation fluence for typical type I and type II NWQDs, i.e., NW heterostructures with and without the internal ZnSe shell, are demonstrated, respectively. It is found that for both cases additional emission lines appear at higher energy with increasing excitation fluence. Considering the result from the type II NWQD shown in Figure 5b, it can be seen that in the low excitation regime, the optical emission spectrum consists of only one emission line at 2.083 eV. Its intensity increases almost linearly with increasing excitation fluence. Then, an additional line appears at 2.109 eV and its intensity increases super-linearly with the exponent twice larger than for the first line. This behavior leads us to associate the line at 2.083 eV to the single exciton, X, and the second line at 2.109 eV to the biexciton emission, XX. Finally, a third emission line appears at 2.128 eV and its intensity increases with an exponent of 2.1, i.e., higher than in the case of XX emission. That is why we speculate that it might be related to multiexcitonic complexes involving more charge carriers than XX, such as e.g., triexcitons or charged biexcitons.

A quite similar behavior is observed in the case of the type I NWQD presented in Figure 5a. There are, however, two important differences. Firstly, only X and XX lines are observed and the third line does not appear. This might be explained by the fact that in type I QDs the carrier lifetime is relatively short which may impede the formation of multiexcitonic complexes with a large number of charge carriers. Secondly, the biexciton binding energy defined as the X-XX spectral distance is distinctly smaller than measured in type II NWQDs. In order to verify the latter point a statistics of biexciton binding energies in type I and type II NWQDs has been performed, Figure 5c. The striking result is that, indeed, the absolute values of XX-binding energy for type II NWQDs, i.e., with an internal ZnSe shell, are significantly larger than in type I NWQDs, whereas the largest values can reach even 50 meV. This observation can be explained be the fact that the biexciton binding energy depends remarkably on the electron-hole spatial separation[21,35], which is obviously larger in the case of type II heterostructures. Responsible for this effect is the repulsive Coulomb interaction between single-sign charge carriers which becomes important when electrons and holes are spatially separated.

At this stage it is important to note that the observation of XX complexes in the emission spectra from the axial NW insertions indicates that the carriers are, indeed, confined in zero dimensional structures. In the emission spectra from structures with a higher dimensionality this kind of emission cannot be detected. For instance, the PL-lines coming from individual $Zn_{0.9}Mg_{0.1}Te$ NW-cores do not show any XX emission. With an increasing excitation fluence their intensity increases linearly, its spectral position shifts typically towards the lower energies and the spectral width increases as an effect of the sample heating (not shown).

Conclusions

NWQDs composed of a short $Zn_{0.97}Mg_{0.03}Te$ low bandgap axial insertion within $Zn_{0.9}Mg_{0.1}Te$ NW-cores are grown by molecular beam epitaxy by employing the vapor liquid solid growth mechanism. The optical emission from these nanostructures has been investigated by means of PL, TRPL, CL and μ-PL-measurements. In particular, CL maps performed with high spatial resolution allow for identification of the emission lines related to NWQDs and NW-cores. The presence of XX emission in the μPL spectrum from individual NWQDs indicates the zero dimensional confinement of carriers in these structures. Coating the NW with a thin ZnSe shell significantly impacts the optical properties of NWQDs. First of all, the emission energy exhibits a significant redshift from 2.36 eV down to 2.10 eV. Simultaneously, the emission intensity decreases, and the PL decay time increases by one order of magnitude. These effects are the fingerprints of the electron hole spatial separation which takes place in these structures due to the type II band alignment at the $ZnSe/Zn_{0.97}Mg_{0.03}Te$ interface. Finally, μ-PL from individual NWs reveals that the XX emission lines are present also in type II NWQDs. However, the XX binding energy is significantly larger compared to the type I counterparts reaching values up to 50 meV. The latter effect can be explained in terms of the spatial electron-hole separation and is directly related to the increasing impact of the Coulomb repulsion between single-sign carriers in type II heterostructures.

Controlling separately the electrons and holes wavefunctions within a single NW heterostructure may result in novel functionalities which could be applied in the field of quantum information technologies, photodetection and photovoltaics. In particular, the type II NWQDs with a radial electron-hole separation have a great potential to be used in quantum devices employing the excitonic Aharonov-Bohm effect. The latter effect has been recently observed in III-V semiconductor based NW heterostructure[31] and might open novel opportunities in the frame of the quantum information storage[25,26].

**Acknowledgements**

This work has been partially supported by the National Centre of Science (Poland) through grant 2017/26/E/ST3/00253, and by the Foundation for Polish Science through the IRA Programme co-financed by EU within SG OP (Grant No. MAB/2017/1).